\documentclass[%
 reprint,
 amsmath,amssymb,
 aps,
]{revtex4-2}

\usepackage{graphicx}
\usepackage{dcolumn}
\usepackage{amsmath}
\usepackage{booktabs}
\usepackage{bm}
\usepackage{hyperref}
\usepackage[mathlines]{lineno}

\newcommand{\Tr}{\mathrm{Tr}}

\begin{document}

\preprint{APS}

\title{Self-Interacting Gravitational Exciton Condensates \\ from Extra-Dimensional Stabilization}

\author{\large Swapnil Kumar Singh{}${}^{1}$}
\email{swapnilsingh.ph@gmail.com}
\thanks{Corresponding author}

\affiliation{
    ${}^{1}$B.M.S. College of Engineering, \\
    Bangalore, Karnataka, 560019, India
}

\date{\today}

\begin{abstract}
We present a comprehensive study of gravitational exciton dynamics arising from higher-dimensional theories, with a focus on establishing a robust effective framework that incorporates self-interactions, higher-derivative corrections, and quantum effects. Starting from a $D$-dimensional Einstein--Hilbert action on a warped product manifold, we perform a systematic dimensional reduction to obtain a four-dimensional effective action in the Einstein frame. Our formulation extends the classical potential by including self-interacting moduli fields and RG-improved parameters via the Coleman--Weinberg mechanism, thereby accounting for both matter and graviton loop corrections. The resulting renormalization group flow modifies the effective mass and coupling constants, which in turn plays a critical role in moduli stabilization and the low-energy phenomenology.

Taking the non-relativistic limit, we derive a Gross--Pitaevskii equation that governs the dynamics of the gravitational exciton condensate, and couple it self-consistently with Poisson’s equation to capture gravitational backreaction. Through both variational and numerical analyses, we obtain stationary solutions in spherically symmetric, rotating, and anisotropic configurations, and perform a linear stability analysis using the Bogoliubov--de Gennes formalism. Our results reveal a Bogoliubov dispersion relation that exhibits a phonon-like linear regime at low momenta, transitioning to quadratic free-particle behavior at high momenta. 
\end{abstract}


\maketitle

\section{Introduction}

The discovery of the accelerated expansion of the Universe \cite{Riess1998,Perlmutter1999} and the overwhelming evidence for dark matter from observations such as the cosmic microwave background \cite{Planck2018} and galaxy rotation curves \cite{RubinFord1970} have profoundly reshaped modern cosmology. These observations indicate that over 95\% of the energy content of the Universe is composed of dark energy and dark matter, phenomena that remain elusive within the Standard Model of particle physics.

Higher-dimensional theories—motivated by string theory, M–theory, and brane–world scenarios \cite{AppelquistChodosFreund1987,OverduinWesson1997,RandallSundrum1999}—offer a compelling framework for unifying gravity with the other fundamental forces. In these models, the geometry and dynamics of compact extra dimensions are crucial in determining the effective four–dimensional physics. Dimensional reduction leads naturally to moduli fields, whose fluctuations yield massive scalar excitations (often referred to as \emph{gravitational excitons} or \emph{radions}) \cite{Gu_nther_2001}. Their effective potential,
\begin{equation}
\label{eq:Ueff_intro}
U_{\mathrm{eff}}(\varphi)= \Lambda_{\mathrm{eff}} + \frac{1}{2}\,m^2\,\varphi^2 + \frac{\lambda}{4}\,\varphi^4 + \cdots\,,
\end{equation}
emerges from a combination of classical geometric effects, flux stabilization mechanisms \cite{GoldbergerWise1999}, and quantum corrections \cite{Wetterich1988,Maartens2010,GasperiniVeneziano2003}. Due to their naturally suppressed couplings to Standard Model particles \cite{BraxDavis2002,Csaki2000}, these excitons are promising dark matter candidates.

In this work, we extend the conventional effective action by incorporating quartic self–interaction terms as well as higher–derivative corrections, thereby capturing the full nonlinear dynamics of gravitational excitons. This extension is particularly important since it permits the formation of a Bose–Einstein condensate (BEC) of gravitational excitons—a state that could radically alter our understanding of dark matter. The non–relativistic dynamics of such a self–gravitating condensate is effectively described by a Gross–Pitaevskii equation coupled to a Poisson equation \cite{Gross1961,Pitaevskii1961,Bohmer2007,202406.1082}. Furthermore, renormalization group (RG) improvement techniques \cite{ColemanWeinberg1973,Donoghue1994,Burgess2004} are employed to incorporate quantum corrections that stabilize the moduli and generate non–trivial potential landscapes.

Our analysis spans several key aspects of the theory. First, by deriving the extended effective action through dimensional reduction and conformal rescaling, we explain the connection between extra–dimensional dynamics and four–dimensional phenomenology. Next, we incorporate quantum corrections and study the RG evolution of the effective couplings, thereby demonstrating how nonlinearities emerge beyond the quadratic approximation. The non–relativistic limit of the RG–improved theory is then taken to derive a Gross–Pitaevskii description of the gravitational exciton condensate. Stationary solutions are analyzed using both variational methods and linear stability techniques (including the Bogoliubov–de Gennes formalism), and the collective excitation spectrum is derived via Bogoliubov theory \cite{Bogolyubov1947OnTT,PitaevskiiStringari2003}.

Notably, our findings indicate that the condensate exhibits a diffuse density profile and a unique excitation spectrum, with a crossover from linear (phonon–like) behavior at low momenta to quadratic free–particle dispersion at high momenta. These properties may lead to observable astrophysical signatures, such as modifications in cosmic microwave background anisotropies and distinctive dynamics in dwarf galaxies \cite{Marsh2016,Li2014}.

In the following sections, we detail the derivation of the effective action, the incorporation of quantum corrections, the formulation of the condensate dynamics, and the stability analysis of stationary solutions. Numerical simulations of the condensate, together with computations of its Bogoliubov excitation spectrum, confirm the theoretical predictions and highlight promising avenues for future observational tests.

\section{Extended Effective Action with Higher-Derivative and Self-Interacting Terms}
\setcounter{equation}{0}

In this section, we develop an extended formulation of the effective action obtained from a higher-dimensional gravitational theory upon compactification. Our approach incorporates quantum corrections, non-trivial topological effects, flux stabilization, and higher-derivative interactions that naturally arise in string-inspired and extra-dimensional models. Our treatment builds upon and extends the methods described in \cite{Gu_nther_2001,AppelquistChodosFreund1987,guenther1999observableeffectsextradimensions} and further refines the framework discussed in \cite{Cicoli2008,Burgess2003}.

\subsection*{Higher-Dimensional Action and Geometrical Setup}

We begin with the $D$-dimensional Einstein--Hilbert action including a bulk cosmological constant and matter contributions,
\begin{equation}
\label{eq:actionD}
S_D = \frac{1}{2\kappa_D^2}\int_{M} d^Dx\,\sqrt{|g|}\,\Bigl\{ R[g] - 2\Lambda \Bigr\} + S_{\mathrm{matter}},
\end{equation}
where $\kappa_D$ is the gravitational coupling in $D$ dimensions and $\Lambda$ is the cosmological constant. We assume that the underlying manifold factorizes as
\begin{equation}
\label{eq:manifold}
M = M_0 \times M_1 \times \cdots \times M_n,
\end{equation}
with the external spacetime $M_0$ being $(D_0 \equiv d_0+1)$-dimensional (with coordinates $x^\mu$, $\mu=0,1,\dots,d_0$), and each internal space $M_i$ having dimension $d_i$. Furthermore, every internal metric $g^{(i)}_{mn}$ is assumed to satisfy the Einstein condition:
\begin{equation}
\label{eq:Einstein_internal}
R_{mn}[g^{(i)}] = \lambda^i\,g^{(i)}_{mn},\quad m,n=1,\dots,d_i,
\end{equation}
with constant eigenvalues $\lambda^i$. This condition ensures the consistency of the compactification procedure and has been widely employed (see, e.g., \cite{Quevedo1996,ScherkSchwarz1979}).

\subsection*{Warped-Product Metric Ansatz and Moduli Fields}

To capture the dynamics of the extra dimensions, we adopt a warped-product metric ansatz:
\begin{equation}
\label{eq:metric_ansatz}
ds^2 = g^{(0)}_{\mu\nu}(x)\,dx^\mu dx^\nu + \sum_{i=1}^{n} e^{2\beta^i(x)}\,g^{(i)}_{mn}(y_i)\,dy_i^m dy_i^n.
\end{equation}
The warp factors $e^{\beta^i(x)}$ represent the dynamical sizes of the internal spaces and are interpreted as moduli fields. In order to study small fluctuations about a stabilized configuration, we define shifted moduli fields:
\begin{equation}
\label{eq:beta_shift}
\tilde{\beta}^i(x) = \beta^i(x) - \beta_0^i,
\end{equation}
where $\beta_0^i$ denote the stabilized (or present-day) values. These moduli will play a central role in the effective external theory and influence the spectral properties of the fluctuation operator (cf. \cite{Polchinski1998}).

\subsection*{Conformal Rescaling and Dimensional Reduction}

To obtain a $(D_0)$-dimensional effective action in the Einstein frame, we perform a conformal rescaling of the external metric:
\begin{align}
g^{(0)}_{\mu\nu}(x) &= \Omega^2(x)\,\tilde{g}^{(0)}_{\mu\nu}(x), \\
\text{with} \quad \Omega^2(x) &= \exp\left[-\frac{2}{D_0-2} \sum_{i=1}^{n} d_i\,\tilde{\beta}^i(x)\right].
\end{align}

Integration over the internal coordinates yields the internal volume,
\begin{equation}
\label{eq:internal_volume}
V_{D'} = \prod_{i=1}^{n}\int_{M_i} d^{d_i}y\,\sqrt{|g^{(i)}|},
\end{equation}
which relates the $D$-dimensional gravitational coupling $\kappa_D$ to its $(D_0)$-dimensional counterpart,
\begin{equation}
\label{eq:kappa0}
\kappa_0^2 = \frac{\kappa_D^2}{V_{D'}}.
\end{equation}
Thus, the effective action takes the preliminary form
\begin{align}
\label{eq:action_eff}
S_{\mathrm{eff}} = & \frac{1}{2\kappa_0^2} \int_{M_0} d^{D_0}x\,\sqrt{|\tilde{g}^{(0)}|} \Biggl\{ \tilde{R}[\tilde{g}^{(0)}] \nonumber \\
& - \bar{G}_{ij}\,\tilde{g}^{(0)\mu\nu}\,\partial_\mu\tilde{\beta}^i\,\partial_\nu\tilde{\beta}^j 
- 2\,U_{\mathrm{eff}}(\tilde{\beta}) \Biggr\}.
\end{align}

where the moduli space metric is given by
\begin{equation}
\label{eq:moduli_metric}
\bar{G}_{ij} = d_i\,\delta_{ij} + \frac{d_i\,d_j}{D_0-2}.
\end{equation}

\subsection*{Effective Potential and Loop Corrections}

The effective potential in Eq.~\eqref{eq:action_eff} arises from both classical curvature contributions and quantum effects. Explicitly, we have
\begin{align}
U_{\mathrm{eff}}(\tilde{\beta}) &= \exp\left[-\frac{2}{D_0 - 2} \sum_{i=1}^{n} d_i \, \tilde{\beta}^i(x)\right] \\
&\quad \times \left( -\frac{1}{2} \sum_{i=1}^{n} \tilde{R}_i \, e^{-2 \tilde{\beta}^i(x)} + \Lambda \right)
\end{align}

with $\tilde{R}_i = R_i\,e^{-2\beta_0^i}$. In addition, one-loop quantum corrections from heavy Kaluza--Klein modes contribute to the potential through the functional determinant
\begin{equation}
\label{eq:1loop}
V_{1\text{-loop}}(\varphi) = \frac{1}{2}\,\Tr\log\!\Bigl(-\Box + m^2(\varphi)\Bigr),
\end{equation}
which, upon employing zeta-function regularization and the heat kernel expansion \cite{Tseytlin1987,BarvinskyVilkovisky1985,BirrellDavies1982}, can be expressed as
\begin{equation}
\label{eq:heatkernel}
V_{1\text{-loop}}(\varphi) = \frac{1}{2(4\pi)^{D_0/2}} \sum_{k=0}^\infty a_k(\varphi) \,\Gamma\!\Bigl(k-\frac{D_0}{2}\Bigr) \Bigl[m^2(\varphi)\Bigr]^{\frac{D_0}{2}-k}.
\end{equation}
This expression, which also determines the renormalization group flow of the effective couplings \cite{JackOsborn1984}, necessitates analytic continuation for odd $D_0$.

\subsection*{Canonical Normalization in the Single Modulus Case}

To explain the key features of our framework, we now focus on the single modulus case ($n=1$). In this case, we define
\begin{equation}
\label{eq:single_modulus}
\tilde{\beta} \equiv \tilde{\beta}^1,\quad d\equiv d_1,\quad \tilde{R}\equiv \tilde{R}_1.
\end{equation}
We introduce a canonically normalized scalar field $\varphi(x)$ via
\begin{equation}
\label{eq:canonical_field}
\varphi(x) = Q\,\tilde{\beta}(x),
\end{equation}
with the normalization constant $Q$ determined by the condition
\begin{equation}
\label{eq:canonical_kinetic}
Q^2\,\bar{G} = 1, \quad \text{with} \quad \bar{G} = d + \frac{d^2}{D_0-2}.
\end{equation}

\subsection*{Perturbative Expansion, Self-Interactions, and Higher-Derivative Corrections}

Assuming that the effective potential is stabilized at $\tilde{\beta}=0$ (or equivalently $\varphi=0$), we perform a Taylor expansion around the minimum:
\begin{align}
U_{\mathrm{eff}}(\varphi) &= U_{\mathrm{eff}}(0) + \frac{1}{2} U_{\mathrm{eff}}''(0) \varphi^2 + \frac{1}{3!} U_{\mathrm{eff}}^{(3)}(0) \varphi^3 \nonumber \\
&\quad + \frac{1}{4!} U_{\mathrm{eff}}^{(4)}(0) \varphi^4 + \cdots.
\end{align}
Motivated by quantum corrections and the integration of heavy modes \cite{Cicoli2008,Burgess2003}, we parameterize the potential as

\begin{align}
U_{\mathrm{eff}}(\varphi) &= \Lambda_{\mathrm{eff}} + \frac{1}{2} m^2 \varphi^2 + \frac{\lambda}{4} \varphi^4 \nonumber \\
&\quad + \frac{\lambda_6}{6!} \varphi^6 + \frac{\lambda_8}{8!} \varphi^8 + \mathcal{O}(\varphi^{10})
\end{align}

where $\Lambda_{\mathrm{eff}} = U_{\mathrm{eff}}(0)$ and $m^2 = U_{\mathrm{eff}}''(0)$, while the higher-order couplings encode further stringy and quantum corrections.

In addition to the standard kinetic term, extra-dimensional effective theories naturally include higher-derivative corrections. The leading correction is given by
\begin{equation}
\label{eq:higher_deriv}
\Delta\mathcal{L}_{\mathrm{deriv}} = \alpha\, \Bigl(\tilde{g}^{(0)\mu\nu}\,\partial_\mu\varphi\,\partial_\nu\varphi\Bigr)^2,
\end{equation}
with $\alpha$ a coupling constant of appropriate mass dimension. Such corrections affect the ultraviolet behavior of the theory and may modify scattering amplitudes and vacuum stability \cite{WeinbergBook,Donoghue1994}.

\paragraph{Spectral Analysis of the Fluctuation Operator.}  
To study the impact of higher-derivative terms on the spectrum of moduli fluctuations, consider the eigenvalue equation for the modified operator:
\begin{equation}
\label{eq:eigenvalue}
\mathcal{O}\,\delta\varphi = \lambda\,\delta\varphi, \quad \text{with} \quad \mathcal{O} = -\Box + m^2 + \xi\,\tilde{R}[\tilde{g}^{(0)}] + \alpha\,\Box^2.
\end{equation}
Employing a plane-wave ansatz $\delta\varphi \sim \exp(i k_\mu x^\mu)$, the dispersion relation becomes
\begin{equation}
\label{eq:dispersion}
\alpha\, (k^2)^2 + k^2 + m^2 + \xi\,\tilde{R} = 0.
\end{equation}
The quartic nature of this equation in $k^2$ requires the use of methods akin to Cardano's approach for depressed quartic equations \cite{Brax2004}, revealing branch cuts in the propagator and indicating potential instabilities unless $\alpha$ is sufficiently suppressed. This analysis mirrors the nonlocal operator techniques discussed in \cite{Polchinski1998}.

\subsection*{Non-Minimal Couplings and Interactions with Matter}

The interplay between curvature and moduli dynamics is incorporated via a non-minimal coupling between the scalar field and the Ricci scalar:
\begin{equation}
\label{eq:nonminimal_coupling}
\Delta\mathcal{L}_{\mathrm{NM}} = \xi\,\tilde{R}[\tilde{g}^{(0)}]\,\varphi^2,
\end{equation}
with $\xi$ a dimensionless parameter. Such couplings naturally emerge in theories with conformal symmetry and have important implications in both early- and late-time cosmology \cite{FaraoniBook,NojiriOdintsov2011}. Additional interactions with matter fields are also included. For example, a Yukawa interaction with fermions is given by
\begin{equation}
\label{eq:Yukawa_interaction}
\Delta\mathcal{L}_{\mathrm{Yukawa}} = g_Y\,\bar{\psi}\psi\,\varphi,
\end{equation}
and a coupling to gauge fields is described by
\begin{equation}
\label{eq:gauge_interaction}
\Delta\mathcal{L}_{\mathrm{gauge}} = \zeta\,\varphi\,F_{\mu\nu}F^{\mu\nu}.
\end{equation}
These interactions not only open novel decay channels but also influence the renormalization group evolution of the effective couplings \cite{ArkaniHamed2002,Dimopoulos2001}.

\subsection*{Flux Stabilization, Topology, and Non-Perturbative Contributions}

If the internal space exhibits non-trivial topology (e.g., orbifolds or Calabi--Yau manifolds) or supports background fluxes \cite{Giddings2002,Kachru2003}, additional contributions to the effective potential arise. Flux-induced superpotentials may introduce exponential or other non-polynomial corrections that aid in moduli stabilization. Non-perturbative effects, such as instanton contributions and vacuum tunneling, further enrich the vacuum structure, potentially leading to metastable vacua of phenomenological interest \cite{Douglas2007}.

\subsection*{Final Extended Effective Action}

Collecting all contributions, the complete effective action for the canonically normalized scalar field $\varphi$ in $D_0$ dimensions is given by
\begin{align}
\label{eq:final_extended_action}
S_{\mathrm{eff}} &= \frac{1}{2\kappa_0^2}\int_{M_0} d^{D_0}x\,\sqrt{|\tilde{g}^{(0)}|}\,\Biggl\{ \tilde{R}[\tilde{g}^{(0)}] - 2\Lambda_{\mathrm{eff}} - \xi\,\tilde{R}[\tilde{g}^{(0)}]\,\varphi^2 \Biggr\} \nonumber\\[1ex]
&\quad - \frac{1}{2}\int_{M_0} d^{D_0}x\,\sqrt{|\tilde{g}^{(0)}|}\,\Biggl\{ \tilde{g}^{(0)\mu\nu}\,\partial_\mu\varphi\,\partial_\nu\varphi + m^2\,\varphi^2 + \lambda\,\varphi^4 \nonumber\\[1ex]
&\quad\qquad\quad + \lambda_6\,\varphi^6 + \lambda_8\,\varphi^8 + \alpha\,\Bigl(\tilde{g}^{(0)\mu\nu}\,\partial_\mu\varphi\,\partial_\nu\varphi\Bigr)^2 + \cdots \Biggr\} \nonumber\\[1ex]
&\quad + S_{\mathrm{int}},
\end{align}
where $S_{\mathrm{int}}$ collects additional interaction terms (such as the Yukawa and gauge couplings given in Eqs.~\eqref{eq:Yukawa_interaction} and \eqref{eq:gauge_interaction}), as well as further corrections from fluxes, topological effects, and higher-loop contributions.

This extended effective action forms the foundation for exploring a broad spectrum of phenomena, including gravitational exciton dynamics, moduli stabilization, inflationary scenarios, and late-time cosmic acceleration. Notably, the inclusion of higher-order self-interactions and derivative corrections permits the possibility of Bose--Einstein condensates (BECs) of gravitational excitations, with potential observational signatures \cite{Marsh2016}.
 
\medskip

\begin{figure}[ht]
  \centering
  \includegraphics[width=0.5\textwidth]{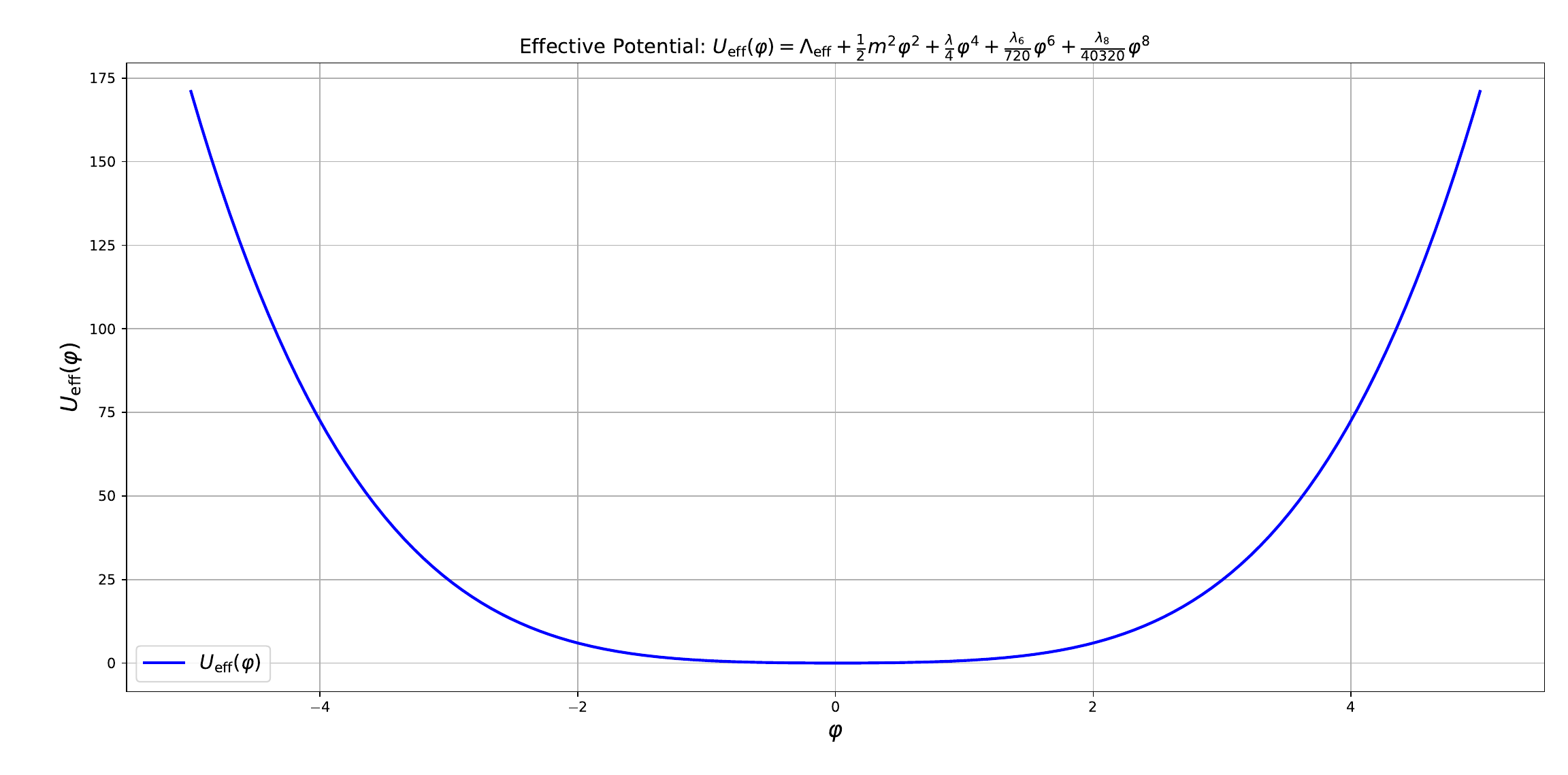}
  \caption{Effective potential $U_{\mathrm{eff}}(\varphi)$ as a function of the canonically normalized field $\varphi$, for $\Lambda_{\mathrm{eff}}=0$, $m^2=1$, $\lambda=1$, $\lambda_6=0.1$, and $\lambda_8=0.01$. The inclusion of higher-order terms results in a multi-branched vacuum structure, hinting at possible phase transitions.}
  \label{fig:potential_extended}
\end{figure}

Figure~\ref{fig:potential_extended} displays a representative effective potential,
\[
U_{\mathrm{eff}}(\varphi) = \Lambda_{\mathrm{eff}} + \frac{1}{2}\,m^2\,\varphi^2 + \frac{\lambda}{4}\,\varphi^4 + \frac{\lambda_6}{6!}\,\varphi^6 + \frac{\lambda_8}{8!}\,\varphi^8,
\]
for the parameter choices
\[
\Lambda_{\mathrm{eff}}=0,\quad m^2=1,\quad \lambda=1,\quad \lambda_6=0.1,\quad \lambda_8=0.01.
\]
The presence of sextic and octic terms yields a rich vacuum structure that may facilitate nontrivial phase transitions and metastable states.

\section{Quantum Corrections and Renormalization Group Effects}
\setcounter{equation}{0}

Having established the classical extended effective action, we now incorporate quantum corrections within the framework of effective field theory. Our analysis employs the Coleman--Weinberg mechanism alongside gravitational loop effects to derive one-loop corrections and study the ensuing renormalization group (RG) evolution of both self-couplings and the moduli parameters controlling the compactification scale \cite{ColemanWeinberg1973,Donoghue1994,Burgess2004}.

\subsection*{One-Loop Effective Potential: Matter and Graviton Contributions}

Starting from the classical effective potential,
\begin{equation}
\label{eq:UeffClassical_exp}
U_{\mathrm{eff}}^{\mathrm{cl}}(\varphi) = \Lambda_{\mathrm{eff}} + \frac{1}{2}\,m^2\,\varphi^2 + \frac{\lambda}{4}\,\varphi^4,
\end{equation}
quantum corrections are obtained by evaluating the one-loop functional determinant. For a scalar field fluctuating about the background $\varphi$, the matter loop contribution is given by the Coleman--Weinberg formula:
\begin{equation}
\label{eq:1loop_V_exp}
V_{\rm 1-loop}^{\rm matter}(\varphi) = \frac{1}{64\pi^2}\,\Bigl[M^2(\varphi)\Bigr]^2 \left[ \ln\!\left( \frac{M^2(\varphi)}{\mu^2} \right) - \frac{3}{2} \right],
\end{equation}
with the field-dependent mass-squared defined as
\begin{equation}
\label{eq:M2phi}
M^2(\varphi) = m^2 + 3\lambda\,\varphi^2,
\end{equation}
and $\mu$ representing the renormalization scale. A more detailed derivation follows from the background field method. In dimensional regularization the functional determinant
\[
\Delta U_{\mathrm{eff}}^{\rm matter}(\varphi) = \frac{1}{2} \int \frac{d^{D_0}k}{(2\pi)^{D_0}} \ln\!\Bigl[k^2+M^2(\varphi)\Bigr],
\]
reduces to Eq.~\eqref{eq:1loop_V_exp} after proper subtraction of divergences \cite{PeskinSchroeder1995,ZinnJustin2002}.

Graviton loops, computed within effective quantum gravity using e.g. the Vilkovisky--DeWitt formalism, contribute additional logarithmic and non-local corrections suppressed by inverse powers of the Planck mass $M_{Pl}$. Schematically, we express the gravitational contribution as
\begin{equation}
\label{eq:1loop_grav}
V_{\rm 1-loop}^{\rm grav}(\varphi) = \frac{1}{64\pi^2}\,f_{\rm grav}(\varphi,\mu,M_{Pl}),
\end{equation}
so that the full one-loop effective potential becomes
\begin{equation}
\label{eq:full_Veff_exp}
\begin{split}
U_{\mathrm{eff}}(\varphi) &= U_{\mathrm{eff}}^{\mathrm{cl}}(\varphi) + \Delta U_{\mathrm{eff}}^{\rm matter}(\varphi) + \Delta U_{\mathrm{eff}}^{\rm grav}(\varphi) \\
&= \Lambda_{\mathrm{eff}} + \frac{1}{2}\,m^2\,\varphi^2 + \frac{\lambda}{4}\,\varphi^4 \\
&\quad + \frac{1}{64\pi^2}\,\Bigl[m^2+3\lambda\,\varphi^2\Bigr]^2 \left[ \ln\!\left( \frac{m^2+3\lambda\,\varphi^2}{\mu^2} \right) - \frac{3}{2} \right] \\
&\quad + \frac{1}{64\pi^2}\,f_{\rm grav}(\varphi,\mu,M_{Pl}).
\end{split}
\end{equation}
Here, the function $f_{\rm grav}(\varphi,\mu,M_{Pl})$ encapsulates the graviton loop corrections computed via methods described in \cite{Reuter1998,Litim2004,Codello2009}.

\subsection*{Emergent Scale Invariance and the Dilaton Mode}

An intriguing feature of the quantum-corrected potential is that, over a suitable range of parameters, the logarithmic terms induce an approximate scale invariance. In this regime the potential becomes nearly flat over an extended region, thereby facilitating spontaneous symmetry breaking and the emergence of a light dilaton mode \cite{Wetterich1988,Shaposhnikov2009,Zumino1970,ZinnJustin2002}. Such a dilaton may address the hierarchy problem and have implications for dark energy phenomenology.

\subsection*{Renormalization Group Flow in a Multidimensional Moduli Space}

Quantum corrections render the couplings scale dependent. The renormalization group evolution is governed by the Callan--Symanzik equation,
\begin{equation}
\label{eq:CallanSymanzik}
\left( \mu \frac{\partial}{\partial \mu} + \beta_{m^2}\,\frac{\partial}{\partial m^2} + \beta_{\lambda}\,\frac{\partial}{\partial \lambda} - \gamma\,\varphi\,\frac{\partial}{\partial \varphi} \right) U_{\mathrm{eff}}(\varphi) = 0,
\end{equation}
with the beta functions for the scalar mass and self-coupling defined as
\begin{equation}
\label{eq:beta_functions}
\beta_{m^2} \equiv \mu\,\frac{d m^2}{d\mu} = \frac{3\lambda}{16\pi^2}\,m^2 + \beta_{m^2}^{\rm grav}, \qquad
\beta_{\lambda} \equiv \mu\,\frac{d\lambda}{d\mu} = \frac{9\lambda^2}{16\pi^2} + \beta_{\lambda}^{\rm grav}.
\end{equation}
Additional gravitational corrections $\beta_{m^2}^{\rm grav}$ and $\beta_{\lambda}^{\rm grav}$ are computed within the effective theory of gravity \cite{Donoghue1994,Burgess2004,PeskinSchroeder1995,MachacekVaughn1984}. In a multidimensional moduli space, the RG flow becomes even richer, with couplings mixing as
\begin{equation}
\label{eq:RG_matrix}
\mu\,\frac{d\vec{g}}{d\mu} = \mathbf{B}(\vec{g}),
\end{equation}
where $\vec{g}=\{m^2,\lambda,\dots\}$ and the matrix $\mathbf{B}$ encodes contributions from both matter and graviton loops. Fixed points and critical behavior in this flow may lead to cascades of phase transitions and are crucial for extra-dimensional stabilization.

\subsection*{Non-Perturbative Phenomena and Vacuum Tunneling}

Beyond perturbation theory, non-perturbative effects such as instanton-induced vacuum tunneling between distinct compactification vacua can alter the effective potential dramatically. In the dilute gas approximation, instanton contributions modify the potential as
\begin{equation}
\label{eq:instanton}
\Delta U_{\mathrm{eff}}^{\rm inst}(\varphi) \sim A(\varphi)\,\exp\!\Bigl[-\frac{S_{\rm inst}(\varphi)}{\hbar}\Bigr],
\end{equation}
where $S_{\rm inst}(\varphi)$ is the instanton action and $A(\varphi)$ is the fluctuation determinant. These non-analytic terms can trigger first-order phase transitions, as originally discussed in \cite{Coleman1977,Callan1977,Lee1977}.

\subsection*{Quantum-Corrected Parameters and Renormalization Conditions}

To quantify the impact of quantum corrections, we expand the effective potential about its minimum at $\varphi=0$ and define the renormalized parameters via
\begin{equation}
\label{eq:renormalized_params}
m_{\rm eff}^2 = U_{\mathrm{eff}}''(0) = m^2 + \Delta m^2, \qquad \lambda_{\rm eff} = \frac{1}{6}\,U_{\mathrm{eff}}^{(4)}(0) = \lambda + \Delta\lambda.
\end{equation}
A straightforward calculation yields, for the mass correction,
\begin{equation}
\label{eq:delta_m2_exp}
\Delta m^2 = \frac{3\lambda}{32\pi^2}\,m^2 \left[ \ln\!\left( \frac{m^2}{\mu^2} \right) - 1 \right] + \Delta m^2_{\rm grav},
\end{equation}
with analogous expressions for $\Delta\lambda$, including both matter and gravitational contributions \cite{Donoghue1994,Burgess2004}. The renormalization conditions guarantee that physical observables remain independent of the arbitrary scale $\mu$ when an RG-improved effective potential is employed.

\subsection*{Numerical Analysis and RG Scale Dependence}

To illustrate the interplay between quantum corrections and RG evolution, we perform a numerical minimization of the effective potential for several renormalization scales. Table~\ref{tab:numerical_results} lists the computed location of the minimum $\varphi_{\mathrm{min}}$, the value of the potential at the minimum $U_{\mathrm{min}}$, and the quantum-corrected mass squared $m_{\rm eff}^2$ for $\mu=0.5$, 1.0, and 2.0.

\begin{table}[ht]
\centering
\begin{tabular}{cccc}
\toprule
$\mu$ & $\varphi_{\mathrm{min}}$ & $U_{\mathrm{min}}$ & $m_{\rm eff}^2$ \\
\midrule
0.50 & 0.00000 & $-1.80012\times10^{-4}$ & 1.00734 \\
1.00 & 0.00000 & $-2.37472\times10^{-3}$ & 0.98100 \\
2.00 & 0.00000 & $-4.56942\times10^{-3}$ & 0.95467 \\
\bottomrule
\end{tabular}
\caption{Numerical values for the effective potential minimum and the quantum-corrected mass squared $m_{\rm eff}^2$ for various renormalization scales $\mu$.}
\label{tab:numerical_results}
\end{table}

\begin{figure}[ht]
\centering
\includegraphics[width=0.5\textwidth]{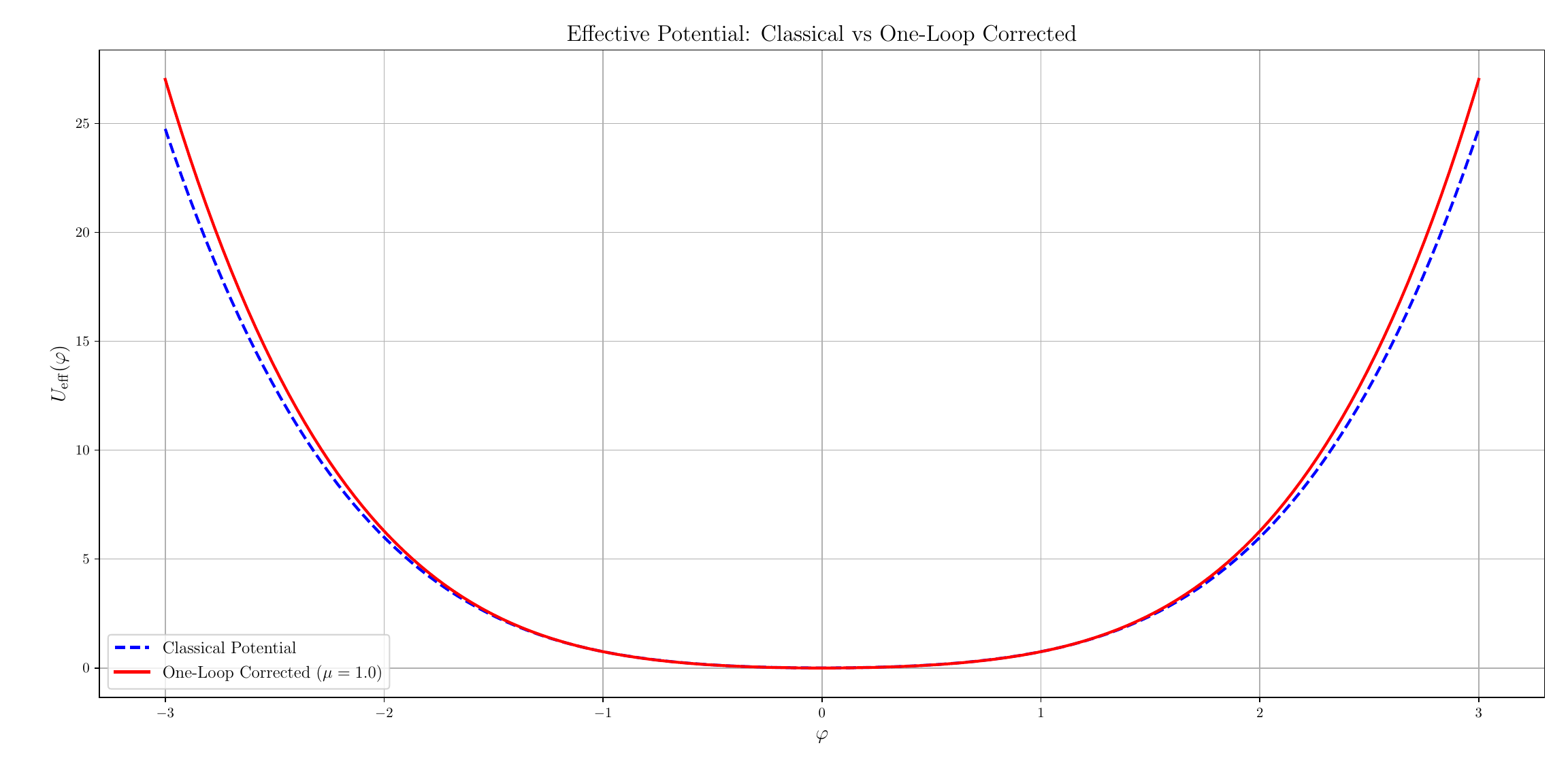}
\caption{Comparison between the classical effective potential and the one-loop corrected potential at $\mu=1.0$. Quantum corrections introduce nontrivial logarithmic structures and modify the curvature near the minimum.}
\label{fig:classical_vs_oneloop}
\end{figure}

\begin{figure}[ht]
\centering
\includegraphics[width=0.5\textwidth]{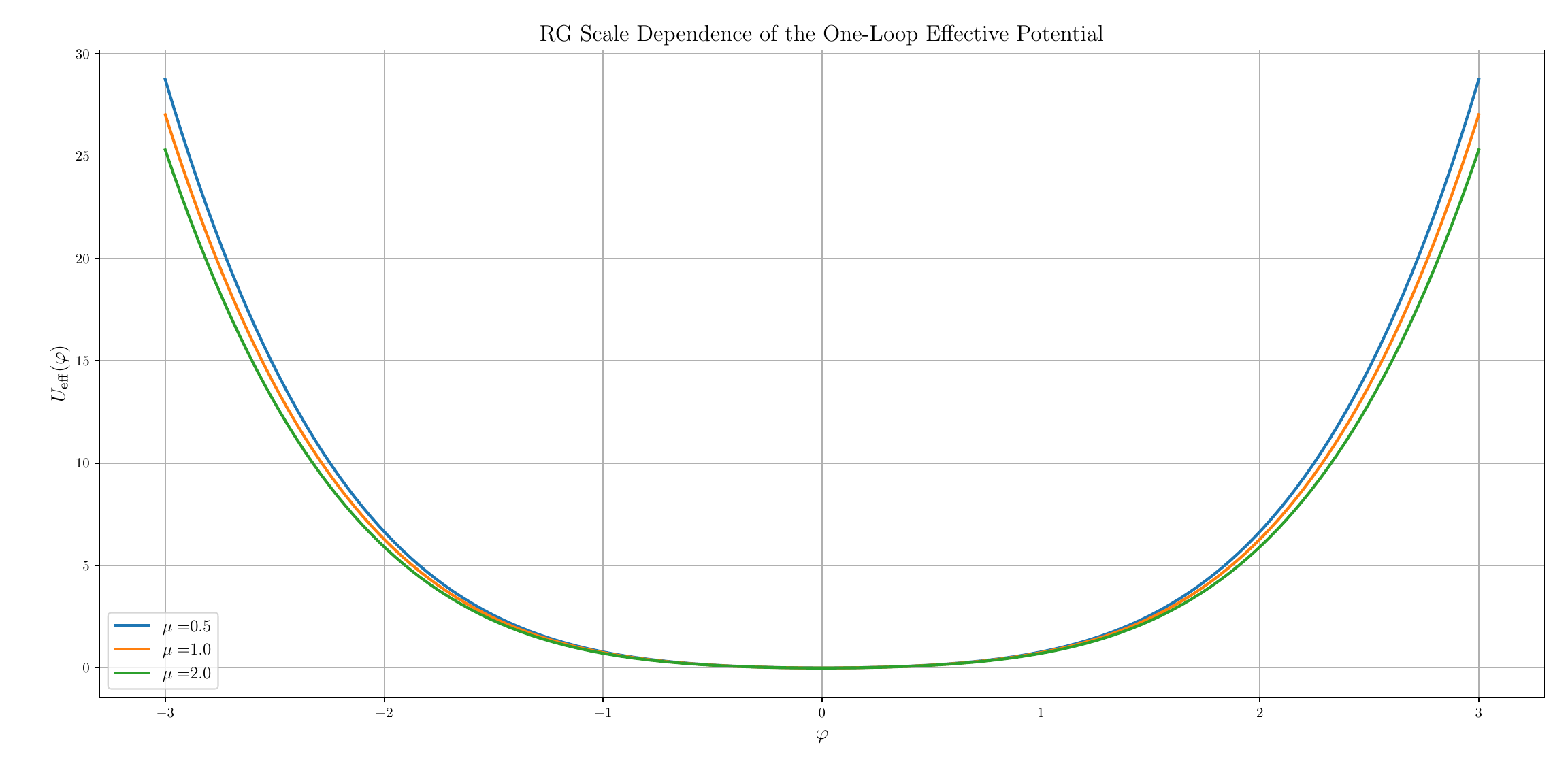}
\caption{RG scale dependence of the one-loop effective potential for renormalization scales $\mu=0.5$, 1.0, and 2.0. The variation in the potential highlights the running of couplings due to quantum corrections.}
\label{fig:rg_scale_dependence}
\end{figure}

Figures~\ref{fig:classical_vs_oneloop} and \ref{fig:rg_scale_dependence} illustrate our findings. In Figure~\ref{fig:classical_vs_oneloop}, we compare the classical potential (Eq.~\eqref{eq:UeffClassical_exp}) with the one-loop corrected potential (Eq.~\eqref{eq:full_Veff_exp}) at $\mu=1.0$. Figure~\ref{fig:rg_scale_dependence} displays the evolution of the effective potential as the renormalization scale is varied, thus reflecting the running of the couplings.

In summary, the interplay between matter and gravitational loop corrections, the emergence of approximate scale invariance, and non-perturbative phenomena enrich the structure of the effective potential. The resulting RG flow, with its fixed points and possible phase transitions, offers valuable insights into exciton dynamics, extra-dimensional stabilization, and potential cosmological implications.

\section{Non-Relativistic Limit and Condensate Dynamics}
\setcounter{equation}{0}

Having derived the RG–improved extended effective action in the relativistic regime, we now focus on its non-relativistic limit and the corresponding condensate dynamics. This limit is particularly relevant for astrophysical applications (e.g., dark matter halos, boson stars) where gravitational exciton modes are expected to condense. For simplicity, we restrict the full theory to four spacetime dimensions ($D_0=4$) and assume a fixed Minkowski background, $\tilde{g}^{(0)}_{\mu\nu}=\eta_{\mu\nu}$, thereby neglecting gravitational backreaction on the condensate dynamics.

\subsection*{Reduction to the Exciton Sector}

In the non-relativistic limit the exciton sector is well described by the action
\begin{equation}
\label{eq:action_phi_NR}
S_\varphi = -\frac{1}{2}\int d^4x\,\left\{ \eta^{\mu\nu}\partial_\mu\varphi\,\partial_\nu\varphi + m_{\rm eff}^2\,\varphi^2 + \lambda_{\rm eff}\,\varphi^4 \right\}\,,
\end{equation}
where $m_{\rm eff}$ and $\lambda_{\rm eff}$ are the quantum–corrected parameters determined from the RG analysis in previous sections. Variation of Eq.~\eqref{eq:action_phi_NR} with respect to $\varphi$ yields a modified Klein–Gordon equation. Expanding the field in Fourier modes and performing a systematic expansion in powers of $|\mathbf{k}|/m_{\rm eff}$ (cf. \cite{Gross1961,Pitaevskii1961}) sets the stage for a non-relativistic reduction.

\subsection*{Non-Relativistic Ansatz and Derivation of the Gross–Pitaevskii Equation}

In the low-energy regime the field $\varphi$ is dominated by its positive-frequency components. We therefore adopt the standard ansatz \cite{PitaevskiiStringari2003}:
\begin{equation}
\label{eq:NR_ansatz}
\varphi(\mathbf{x},t) = \frac{1}{\sqrt{2m_{\rm eff}}}\Bigl[ e^{-im_{\rm eff}t/\hbar}\,\psi(\mathbf{x},t) + e^{im_{\rm eff}t/\hbar}\,\psi^*(\mathbf{x},t) \Bigr]\,,
\end{equation}
where $\psi(\mathbf{x},t)$ is a slowly varying complex amplitude relative to the rapid oscillations encoded in $e^{\pm im_{\rm eff}t/\hbar}$. Substituting Eq.~\eqref{eq:NR_ansatz} into the modified Klein–Gordon equation and neglecting second-order time derivatives of $\psi$ (i.e. assuming $\partial_t\psi \ll m_{\rm eff}\,\psi$) leads, after averaging over the fast oscillations, to the Gross–Pitaevskii (GP) equation:
\begin{equation}
\label{eq:GP_eq}
i\hbar\,\partial_t\psi(\mathbf{x},t) = \left[-\frac{\hbar^2}{2m_{\rm eff}}\nabla^2 + \lambda_{\mathrm{eff}}^{\rm NR}\,|\psi(\mathbf{x},t)|^2 \right]\psi(\mathbf{x},t)\,,
\end{equation}
with the effective nonlinearity parameter given by
\begin{equation}
\label{eq:lambda_eff_NR}
\lambda_{\mathrm{eff}}^{\rm NR} = \frac{3\lambda_{\rm eff}}{2m_{\rm eff}}\,.
\end{equation}
The numerical prefactor in Eq.~\eqref{eq:lambda_eff_NR} arises from the particular normalization in Eq.~\eqref{eq:NR_ansatz} and the projection of the quartic term onto the slowly varying mode \cite{PethickSmith2008}.

\subsection*{Inclusion of Self-Gravity: The Coupled GP--Poisson System}

In many astrophysical and cosmological settings the condensate is self-gravitating. To account for this, we couple the GP equation to the Poisson equation for the gravitational potential $\Phi(\mathbf{x},t)$. The resulting coupled system is
\begin{align}
\label{eq:GP_gravity}
i\hbar\,\partial_t\psi(\mathbf{x},t) &= \left[-\frac{\hbar^2}{2m_{\rm eff}}\nabla^2 + m_{\rm eff}\,\Phi(\mathbf{x},t) \right. \notag \\
&\quad \left. + \lambda_{\mathrm{eff}}^{\rm NR}\,|\psi(\mathbf{x},t)|^2 \right] \psi(\mathbf{x},t)\,, \\[1ex]
\label{eq:Poisson_NR}
\nabla^2\Phi(\mathbf{x},t) &= 4\pi G\,m_{\rm eff}\,|\psi(\mathbf{x},t)|^2\,.
\end{align}

Here, the condensate density is defined as $\rho(\mathbf{x},t)=m_{\rm eff}|\psi(\mathbf{x},t)|^2$. The GP–Poisson system has been extensively studied in the context of Bose–Einstein condensate (BEC) dark matter and boson star formation (see, e.g., \cite{Chavanis2011,Ruffini1969,Schive2014}).

\subsection*{Finite-Temperature Effects and Phase Transition Dynamics}

Realistic condensates are subject to thermal fluctuations. Finite-temperature field theory techniques \cite{KapustaBook,LeBellacBook} enable the derivation of a temperature–dependent GP equation. Near a critical temperature $T_c$, the condensate vanishes and the system undergoes a second–order phase transition characterized by universal critical exponents \cite{PethickSmith2008}. In our framework, finite-temperature corrections can be incorporated either by adding a stochastic noise term to Eq.~\eqref{eq:GP_eq} (as in \cite{Stoof1999}) or by promoting the effective coupling $\lambda_{\mathrm{eff}}^{\rm NR}$ to a temperature–dependent function. These thermal effects are particularly significant in the early Universe, where they may influence structure formation and the stabilization of extra dimensions \cite{Harko2012}.

\subsection*{Rotating Condensates and Vortex Nucleation}

Angular momentum plays a crucial role in many astrophysical scenarios. In a rotating frame the GP equation acquires an additional term corresponding to the Coriolis force. The modified equation is
\begin{align}
i\hbar \, \partial_t \psi(\mathbf{x}, t) &= \left[-\frac{\hbar^2}{2m_{\rm eff}} \nabla^2 + m_{\rm eff} \, \Phi(\mathbf{x}, t) \right. \nonumber \\
&\quad \left. - \Omega L_z + \lambda_{\mathrm{eff}}^{\rm NR} |\psi(\mathbf{x}, t)|^2 \right] \psi(\mathbf{x}, t) \label{eq:GP_rotating}
\end{align}
where $\Omega$ is the angular velocity and $L_z=-i\hbar\,(x\partial_y-y\partial_x)$ denotes the $z$–component of the angular momentum operator. Rotation can induce the nucleation of quantized vortices and the formation of vortex lattices, phenomena well documented in laboratory BECs \cite{Fetter2009,Anderson2001} and proposed as mechanisms affecting galactic rotation curves and boson star dynamics \cite{Ruffini1969,Chavanis2011}.

\subsection*{Numerical Simulations and Coupling with External Fields}

To investigate the condensate dynamics under realistic conditions, we performed numerical simulations of the coupled Gross-Pitaevskii (GP)–Poisson system, as described by Eqs.~\eqref{eq:GP_gravity} and \eqref{eq:Poisson_NR}. These simulations take into account several factors: finite-temperature effects, modeled by introducing a stochastic noise term in the GP equation; self-gravity, which is implemented through a self-consistent solution of the Poisson equation; and potential rotational effects, incorporated by including the Coriolis term in Eq.~\eqref{eq:GP_rotating}, with the rotation parameter $\Omega$ set to zero unless stated otherwise.

Table~\ref{tab:simulation_observables} summarizes key observables from our simulations, including the maximum density, mean density, and the extremal values of the gravitational potential at representative simulation times $t=10$ and $t=20$.
\begin{table}[ht]
    \centering
    \begin{tabular}{ccccc}
        \toprule
        Time & $\varphi_{\rm max}$ & Mean Density & Min $\Phi$ & Max $\Phi$ \\
        \midrule
        10 & 0.01335 & 0.0025 & $-0.00960$ & 0.00501 \\
        20 & 0.00694 & 0.0025 & $-0.00146$ & 0.00082 \\
        \bottomrule
    \end{tabular}
    \caption{Representative observables from numerical simulations of the coupled GP–Poisson system. The decrease in maximum density and the shallowing of the gravitational potential indicate the condensate's relaxation toward equilibrium while conserving particle number.}
    \label{tab:simulation_observables}
\end{table}

\begin{figure}[ht]
    \centering
    \includegraphics[width=0.5\textwidth]{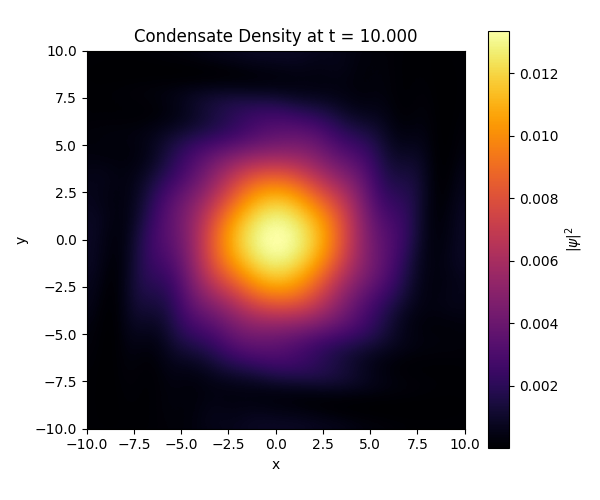}
    \hfill
    \includegraphics[width=0.5\textwidth]{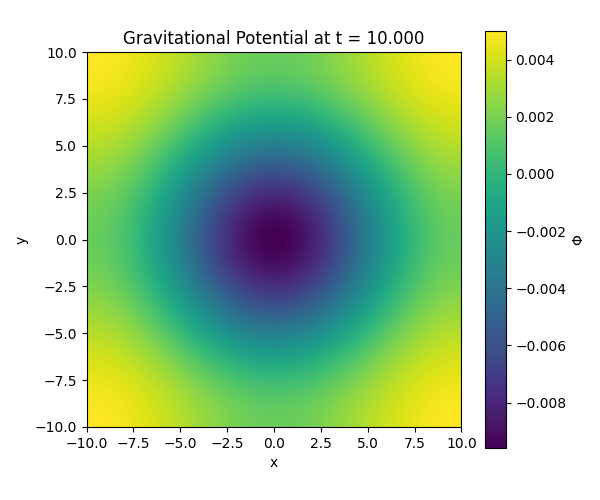}
    \caption{Snapshots at $t=10$: (Left) condensate density exhibiting localized peaks; (Right) the corresponding gravitational potential with deep central wells.}
    \label{fig:t10_snapshot}
\end{figure}

\begin{figure}[ht]
    \centering
    \includegraphics[width=0.5\textwidth]{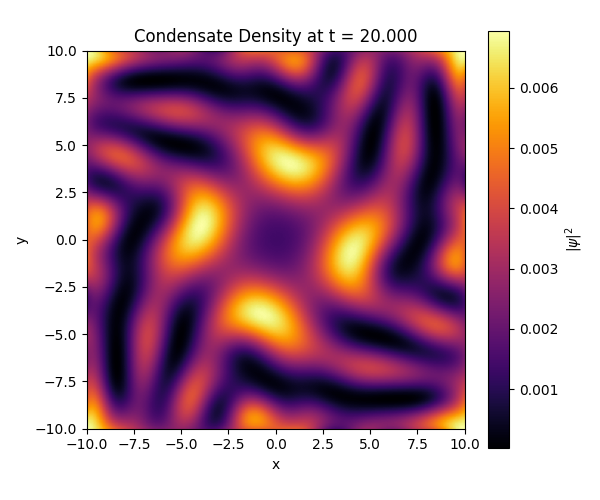}
    \hfill
    \includegraphics[width=0.5\textwidth]{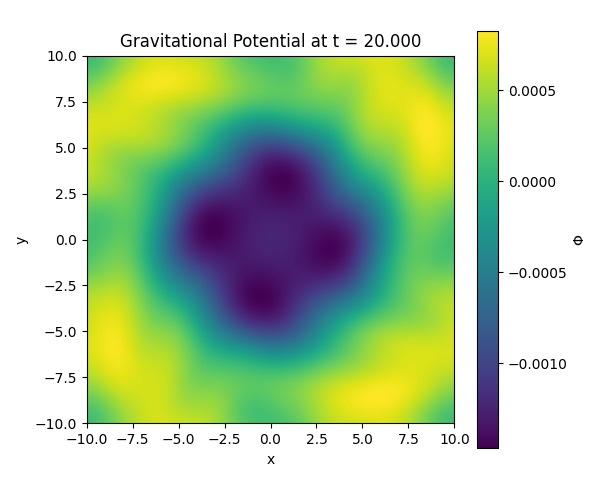}
    \caption{Snapshots at $t=20$: (Left) the density profile becomes smoother; (Right) the gravitational potential is considerably flatter compared to $t=10$.}
    \label{fig:t20_snapshot}
\end{figure}

Figures~\ref{fig:t10_snapshot} and \ref{fig:t20_snapshot} display snapshots of the condensate density and gravitational potential at times $t=10$ and $t=20$, respectively. At $t=10$, pronounced density fluctuations (localized peaks) correlate with deep gravitational potential wells, reflecting the competition between self–gravity and quantum pressure. By $t=20$, these fluctuations merge and dissipate, yielding a smoother density profile and a shallower potential indicative of equilibration.

\medskip
In summary, by taking the non-relativistic limit of the RG–improved effective action, we arrive at a Gross–Pitaevskii description of gravitational exciton condensates. The coupled GP–Poisson system, enriched by finite-temperature corrections and potential rotational effects, offers a robust framework for exploring astrophysical phenomena such as dark matter halo formation, boson star evolution, and vortex dynamics. Future work will incorporate external fields and investigate nontrivial topological defects within this condensate framework.

\section{Stationary Solutions and Stability Analysis}
\setcounter{equation}{0}

To understand the long–term behavior and astrophysical implications of gravitational exciton condensates, it is crucial to study their stationary configurations and analyze their stability. In this section, we investigate both spherically symmetric and more complex solutions (including rotating, anisotropic, and solitonic states), which can model realistic structures such as dark matter halos or boson stars. We further perform a linear stability analysis via the Bogoliubov–de Gennes formalism and discuss the Vakhitov–Kolokolov criterion for stability \cite{Kaup1968,VakhitovKolokolov1973,SeidelSuen1991}.

\subsection*{Stationary Ansatz and Time–Independent Equations}

We begin by seeking stationary solutions of the coupled Gross–Pitaevskii–Poisson system. Assuming a harmonic time dependence,
\begin{equation}
\label{eq:stationary_ansatz}
\psi(\mathbf{x},t) = e^{-i\mu t/\hbar}\,\psi_0(\mathbf{x})\,,
\end{equation}
where $\mu$ is the chemical potential and $\psi_0(\mathbf{x})$ is time–independent. Substituting Eq.~\eqref{eq:stationary_ansatz} into the time–dependent Gross–Pitaevskii equation (cf. Eq.~\eqref{eq:GP_gravity}) yields the time–independent GP equation:
\begin{equation}
\label{eq:TI_GP}
\mu\,\psi_0(\mathbf{x}) = \left[-\frac{\hbar^2}{2m_{\rm eff}}\nabla^2 + m_{\rm eff}\,\Phi(\mathbf{x}) + \lambda_{\mathrm{eff}}^{\rm NR}\,|\psi_0(\mathbf{x})|^2\right]\psi_0(\mathbf{x})\,,
\end{equation}
with the gravitational potential determined self–consistently by
\begin{equation}
\label{eq:TI_Poisson}
\nabla^2\Phi(\mathbf{x}) = 4\pi G\,m_{\rm eff}\,|\psi_0(\mathbf{x})|^2\,.
\end{equation}
The condensate wavefunction is normalized via
\begin{equation}
\label{eq:normalization}
\int d^3x\,|\psi_0(\mathbf{x})|^2 = N\,,
\end{equation}
where $N$ is the total number of excitons. Equations \eqref{eq:TI_GP} and \eqref{eq:TI_Poisson} thus form a nonlinear eigenvalue problem for $\psi_0(\mathbf{x})$ and $\mu$.

\subsection*{Spherically Symmetric Configurations}

For spherically symmetric solutions, we set $\psi_0(\mathbf{x}) = \psi_0(r)$ and $\Phi(\mathbf{x}) = \Phi(r)$, with $r=|\mathbf{x}|$. In spherical coordinates, the Laplacian simplifies to
\begin{equation}
\label{eq:Laplacian_spherical}
\nabla^2 f(r) = \frac{d^2 f}{dr^2} + \frac{2}{r}\frac{df}{dr}\,.
\end{equation}
Thus, the time–independent GP equation reduces to
\begin{align}
\label{eq:GP_spherical}
\mu\,\psi_0(r) &= \left[-\frac{\hbar^2}{2m_{\rm eff}}\left(\frac{d^2}{dr^2} + \frac{2}{r}\frac{d}{dr}\right) \right. \notag \\
&\quad + m_{\rm eff}\,\Phi(r) \notag \\
&\quad \left. + \lambda_{\mathrm{eff}}^{\rm NR}\,|\psi_0(r)|^2\right] \psi_0(r)\,.
\end{align}

and the corresponding Poisson equation becomes
\begin{equation}
\label{eq:Poisson_spherical}
\frac{1}{r^2}\frac{d}{dr}\left(r^2\frac{d\Phi(r)}{dr}\right) = 4\pi G\,m_{\rm eff}\,|\psi_0(r)|^2\,.
\end{equation}
Multiplying Eq.~\eqref{eq:GP_spherical} by $\psi_0^*(r)$ and integrating yields the energy functional of the system. A virial theorem can be derived by scaling arguments \cite{Chavanis2011}, providing further insights into the balance between kinetic, self–interaction, and gravitational energy contributions.

\subsection*{Extensions to Rotating and Anisotropic States}

Many astrophysical systems exhibit non–spherical morphologies. To model such systems, one may generalize the stationary ansatz by incorporating angular momentum:
\begin{equation}
\label{eq:rotating_ansatz}
\psi(\mathbf{x},t) = e^{-i\mu t/\hbar}\,e^{i\ell\theta}\,\psi_0(r,\theta,\phi)\,,
\end{equation}
where $\ell$ denotes the quantized angular momentum. This ansatz modifies the kinetic energy by introducing a centrifugal barrier and, for anisotropic deformations, one may adopt a variational ansatz with anisotropic Gaussian profiles \cite{Fetter2009}. Localized solitonic states (such as Q–balls) can also be obtained by minimizing the energy functional
\begin{equation}
\label{eq:energy_functional}
E[\psi_0] = \int d^3x\,\left[\frac{\hbar^2}{2m_{\rm eff}}|\nabla\psi_0|^2 + \frac{m_{\rm eff}}{2}\,\Phi\,|\psi_0|^2 + \frac{\lambda_{\mathrm{eff}}^{\rm NR}}{2}\,|\psi_0|^4\right]\,,
\end{equation}
subject to the normalization constraint \eqref{eq:normalization}. Variational methods and the Bogoliubov–de Gennes formalism \cite{PitaevskiiStringari2003} can then be employed to map the stability region in parameter space, taking into account the scale dependence induced by RG corrections \cite{Donoghue1994,Burgess2004}.

\subsection*{Linear Stability Analysis}

To assess the dynamical stability of a stationary solution $\psi_0(\mathbf{x})$, we consider small perturbations:
\begin{equation}
\label{eq:perturbation}
\psi(\mathbf{x},t) = e^{-i\mu t/\hbar}\left[\psi_0(\mathbf{x}) + u(\mathbf{x})e^{-i\omega t} + v^*(\mathbf{x})e^{i\omega t}\right]\,.
\end{equation}
Inserting this into the time–dependent GP equation and linearizing in $u$ and $v$, we obtain the Bogoliubov–de Gennes equations. The resulting eigenvalue problem for the excitation frequencies $\omega$ determines the stability of the stationary state. In particular, the Vakhitov–Kolokolov (VK) criterion,
\begin{equation}
\label{eq:VK}
\frac{dN}{d\mu} < 0\,,
\end{equation}
provides a necessary condition for stability \cite{VakhitovKolokolov1973}. A violation of this condition typically signals the onset of collapse or modulational instability \cite{Kaup1968}.

\subsection*{Numerical Stationary Solutions and Convergence}

To obtain explicit stationary solutions, we numerically solve Eqs.~\eqref{eq:GP_spherical} and \eqref{eq:Poisson_spherical} using an imaginary time evolution method on a finite–difference grid in spherical coordinates. In a typical simulation, convergence is achieved after a simulation time of approximately 225.80 (in dimensionless units), yielding a ground state characterized by a chemical potential $\mu \approx 0.05057$ and a total energy $E \approx 0.00587$. The stationary density profile $|\psi_0(r)|^2$ exhibits a maximum of about 0.00336 with a mean density near 0.000564, while the gravitational potential $\Phi(r)$ attains a minimum value of roughly $-0.00588$ at $r=0$ and smoothly approaches zero at large $r$. These results indicate that the condensate relaxes into a diffuse ground state with a central density peak and a shallow potential well.

\begin{figure}[ht]
    \centering
    \includegraphics[width=0.5\textwidth]{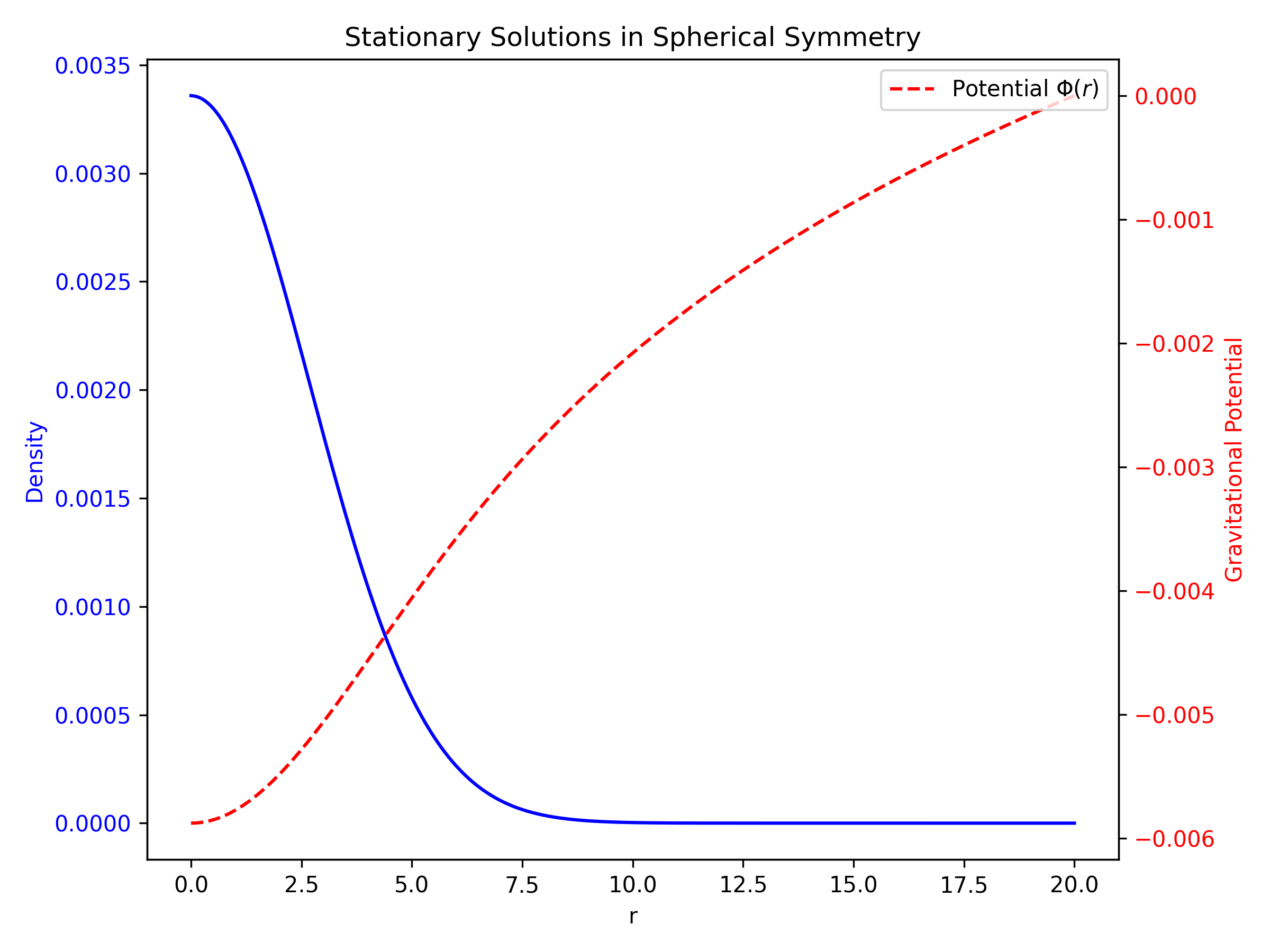}
    \hfill
    \includegraphics[width=0.5\textwidth]{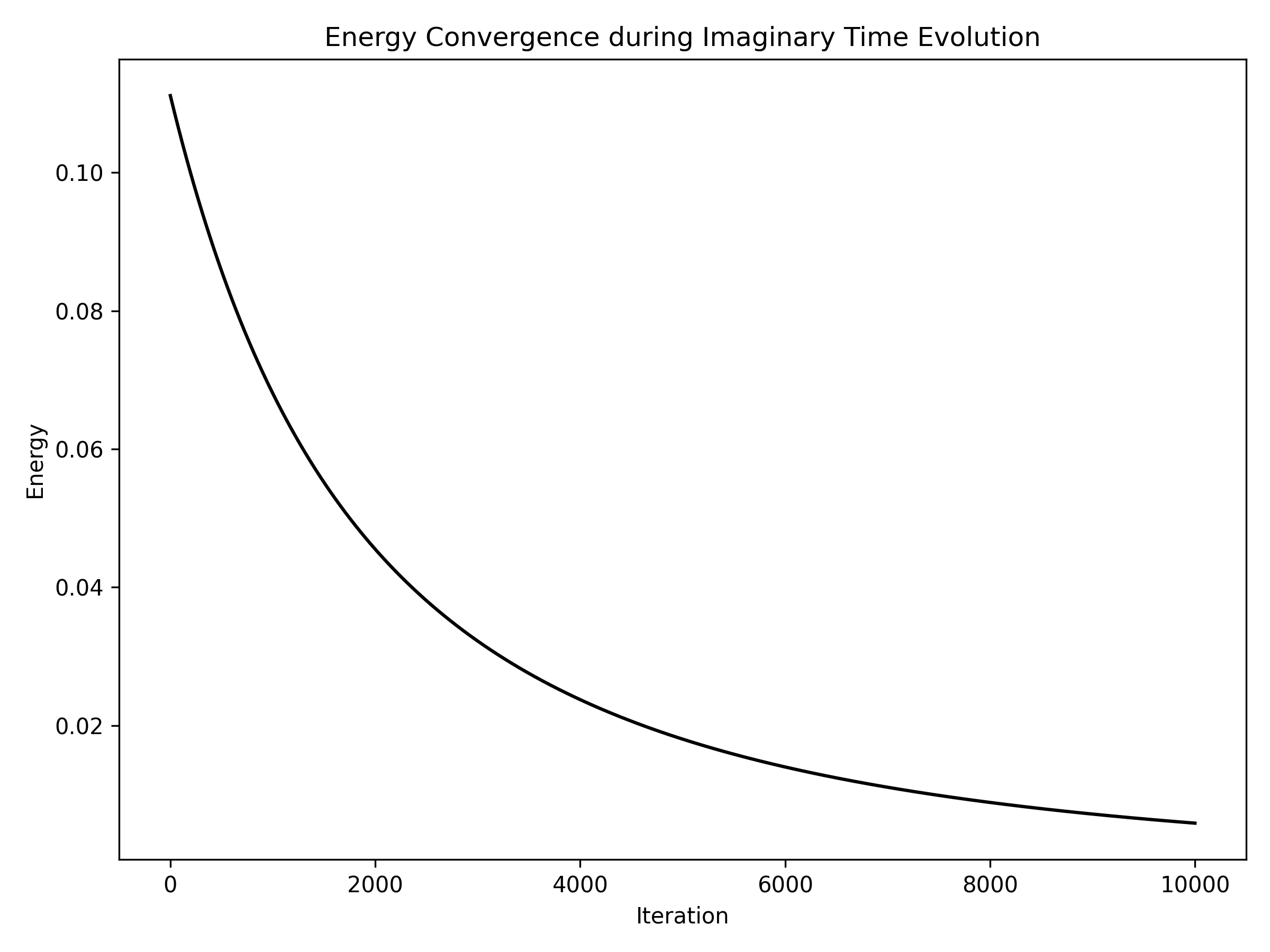}
    \caption{(Left) Converged stationary density $|\psi_0(r)|^2$ (blue solid line) and gravitational potential $\Phi(r)$ (red dashed line) versus radial coordinate $r$. (Right) Convergence of the energy functional during imaginary time evolution, indicating that the system has reached its ground state.}
    \label{fig:stationary_solution}
\end{figure}

Figure~\ref{fig:stationary_solution} (left panel) shows the converged stationary density (blue solid line) and gravitational potential (red dashed line) as functions of $r$, and the right panel illustrates the convergence of the energy functional during the imaginary time evolution, confirming that a stable state is reached.

In summary, our stability analysis demonstrates that gravitational exciton condensates support a rich variety of stationary configurations. The combination of numerical solutions in the spherically symmetric case, extensions to rotating and anisotropic states, and a linear stability analysis based on the Bogoliubov–de Gennes equations provide a robust framework for exploring how RG–improved quantum corrections affect the stabilization of self–gravitating condensates. These results are essential for modeling dark matter structures and boson stars in astrophysical contexts.

\section{Bogoliubov Excitations and Dispersion Relation}
\setcounter{equation}{0}

Collective excitations play a key role in determining the dynamical response and stability of gravitational exciton condensates. In this section, we analyze small fluctuations about a homogeneous condensate state and derive the corresponding Bogoliubov dispersion relation. This analysis not only confirms the presence of phonon–like modes at low momenta but also delineates the crossover to free–particle behavior at higher momenta, thereby providing insight into both the low–energy collective dynamics and high–energy single–particle excitations \cite{PitaevskiiStringari2003,Bogolyubov1947OnTT}.

\subsection*{Linearization about a Homogeneous Condensate}

Assuming that the condensate is spatially homogeneous, we set
\begin{equation}
\label{eq:homogeneous}
\psi_0(\mathbf{x}) = \sqrt{n_0}\,,
\end{equation}
with constant density $n_0$, and neglect any gravitational potential variations over the relevant length scales. To study small perturbations, we write
\begin{equation}
\label{eq:psi_full}
\psi(\mathbf{x},t) = e^{-i\mu t/\hbar}\left[\sqrt{n_0} + \delta\psi(\mathbf{x},t)\right]\,,
\end{equation}
where $\mu$ is the chemical potential. Decomposing the fluctuation into plane–wave modes,
\begin{equation}
\label{eq:fluctuation}
\delta\psi(\mathbf{x},t) = u_{\mathbf{k}}\,e^{i(\mathbf{k}\cdot\mathbf{x}-\omega t)} + v_{\mathbf{k}}^*\,e^{-i(\mathbf{k}\cdot\mathbf{x}-\omega t)}\,,
\end{equation}
and substituting into the time–dependent Gross–Pitaevskii equation \cite{PitaevskiiStringari2003,Bogolyubov1947OnTT}, we retain only linear terms in the fluctuations. This procedure yields the Bogoliubov–de Gennes equations:
\begin{align}
\label{eq:BdG1}
\hbar\omega\,u_{\mathbf{k}} &= \Bigl[\epsilon_k + \lambda_{\mathrm{eff}}^{\rm NR} n_0\Bigr] u_{\mathbf{k}} + \lambda_{\mathrm{eff}}^{\rm NR} n_0\, v_{\mathbf{k}}\,,\\[1ex]
\label{eq:BdG2}
-\hbar\omega\,v_{\mathbf{k}} &= \Bigl[\epsilon_k + \lambda_{\mathrm{eff}}^{\rm NR} n_0\Bigr] v_{\mathbf{k}} + \lambda_{\mathrm{eff}}^{\rm NR} n_0\, u_{\mathbf{k}}\,,
\end{align}
where the free–particle kinetic energy is defined as
\begin{equation}
\label{eq:epsilon_k}
\epsilon_k = \frac{\hbar^2 k^2}{2m_{\rm eff}}\,.
\end{equation}

\subsection*{Derivation of the Bogoliubov Dispersion Relation}

Writing Eqs.~\eqref{eq:BdG1}–\eqref{eq:BdG2} in matrix form,
\begin{equation}
\begin{pmatrix}
\epsilon_k + \lambda_{\mathrm{eff}}^{\rm NR}n_0 & \lambda_{\mathrm{eff}}^{\rm NR}n_0 \\
-\lambda_{\mathrm{eff}}^{\rm NR}n_0 & -\left(\epsilon_k + \lambda_{\mathrm{eff}}^{\rm NR}n_0\right)
\end{pmatrix}
\begin{pmatrix} u_{\mathbf{k}} \\ v_{\mathbf{k}} \end{pmatrix}
= \hbar\omega
\begin{pmatrix} u_{\mathbf{k}} \\ v_{\mathbf{k}} \end{pmatrix}\,,
\end{equation}
non–trivial solutions exist if the determinant of the coefficient matrix vanishes. This condition leads to
\begin{equation}
\label{eq:bog_dispersion}
\hbar^2\omega^2 = \left[\epsilon_k + \lambda_{\mathrm{eff}}^{\rm NR} n_0\right]^2 - \Bigl[\lambda_{\mathrm{eff}}^{\rm NR} n_0\Bigr]^2
= \epsilon_k\Bigl(\epsilon_k+2\lambda_{\mathrm{eff}}^{\rm NR} n_0\Bigr)\,.
\end{equation}
This is the celebrated Bogoliubov dispersion relation originally derived in the context of dilute Bose gases \cite{Bogolyubov1947OnTT}.

\subsection*{Long–Wavelength Limit: Phonon Modes}

In the long–wavelength limit where $\epsilon_k \ll \lambda_{\mathrm{eff}}^{\rm NR} n_0$, Eq.~\eqref{eq:bog_dispersion} simplifies to
\begin{equation}
\label{eq:linear_dispersion}
\hbar\omega \approx \hbar k \sqrt{\frac{\lambda_{\mathrm{eff}}^{\rm NR} n_0}{m_{\rm eff}}}\,.
\end{equation}
This linear dispersion defines the speed of sound in the condensate,
\begin{equation}
\label{eq:sound_speed}
c_s = \sqrt{\frac{\lambda_{\mathrm{eff}}^{\rm NR} n_0}{m_{\rm eff}}}\,,
\end{equation}
so that the low–momentum excitation spectrum is given by
\begin{equation}
\label{eq:phonon_dispersion}
\omega \approx c_s\, k\,.
\end{equation}
These phonon–like modes are essential for understanding superfluid behavior and the low–energy response of the condensate \cite{PethickSmith2008}.

\subsection*{Finite-Temperature and Nonlinear Corrections}

While the derivation above is valid at zero temperature and within the linear regime, finite-temperature effects can lead to damping (via mechanisms such as Landau damping) and modify the dispersion relation by establishing a Landau critical velocity below which excitations remain undamped \cite{PethickSmith2008}. Furthermore, nonlinear corrections—arising from higher–order terms in the fluctuation expansion—can result in the formation of solitons, shock waves, and other coherent structures. Recent studies have also proposed that Bogoliubov excitations might couple to gravitational waves, providing an intriguing observational window into the properties of gravitational exciton condensates \cite{Marsh2016,Li2014}.

\subsection*{Numerical Results and Discussion}

\begin{figure}[ht]
    \centering
    \includegraphics[width=0.5\textwidth]{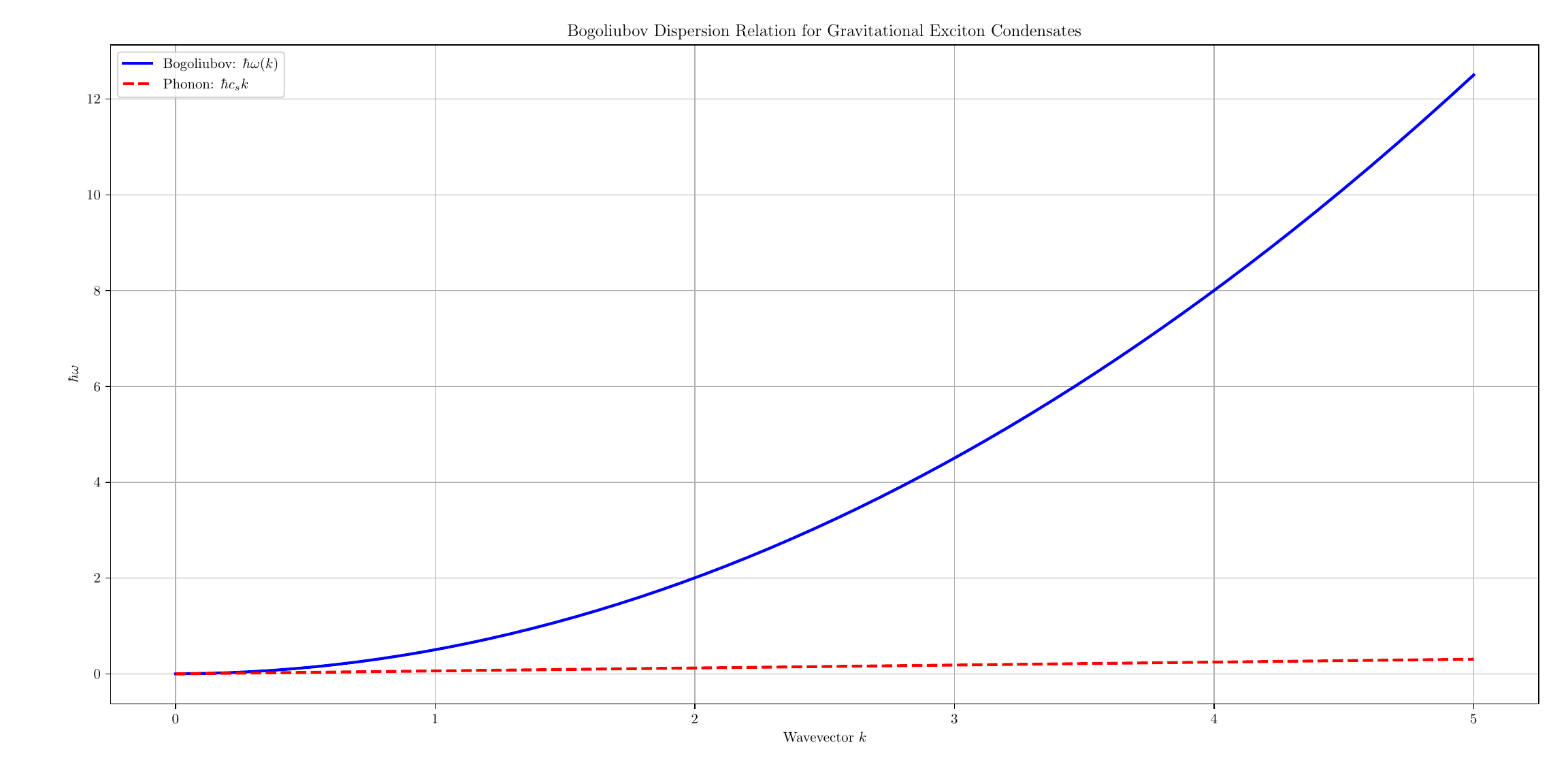}
    \caption{Bogoliubov dispersion relation $\hbar\omega$ as a function of the wavevector $k$, obtained from $\hbar^2\omega^2=\epsilon_k\Bigl(\epsilon_k+2\lambda_{\mathrm{eff}}^{\rm NR}n_0\Bigr)$. The low–$k$ regime exhibits a linear (phonon–like) dispersion with speed of sound $c_s\approx6.1237\times10^{-2}$, while at higher $k$ the dispersion asymptotically approaches the quadratic free–particle limit.}
    \label{fig:dispersion_relation}
\end{figure}

Figure~\ref{fig:dispersion_relation} displays the Bogoliubov dispersion relation $\hbar\omega(k)$ as computed from Eq.~\eqref{eq:bog_dispersion} for representative parameters: $\lambda_{\mathrm{eff}}^{\rm NR} = 1.5000$ and $n_0 = 2.5\times10^{-3}$, yielding a speed of sound $c_s \approx 6.1237\times10^{-2}$. For example, at a low wavevector $k\approx 0.10$, the free–particle energy is $\epsilon_k\approx 5.0335\times10^{-3}$ and the frequency is $\omega\approx 7.9428\times10^{-3}$; at a higher wavevector $k\approx 2.01$, we find $\epsilon_k\approx 2.0134$ and $\omega\approx 2.0171$. These results clearly illustrate the crossover from linear (phonon–like) behavior at low momenta to the quadratic dispersion characteristic of free particles at higher momenta.

In conclusion, the derivation of the Bogoliubov dispersion relation confirms that gravitational exciton condensates support well–defined collective excitations. The emergence of a linear, sound–like regime at low momenta and the crossover to quadratic (free–particle) behavior at higher momenta provide a robust framework for further investigations into finite–temperature effects, nonlinear dynamics, and possible couplings to gravitational phenomena with potential astrophysical signatures.


\section{Conclusion and Discussion}
\setcounter{equation}{0}

In this work, we have developed a comprehensive theoretical framework for gravitational exciton condensates arising from extra-dimensional stabilization. Starting from a higher-dimensional gravitational action, we derived an effective four-dimensional model in which the conformal moduli of the internal spaces manifest as massive scalar fields (gravitational excitons). By extending the effective action to include self-interacting terms—incorporating higher-order and derivative couplings as well as non-minimal curvature interactions—we provided a natural mechanism for both the stabilization of the extra dimensions and the generation of an effective cosmological constant.

Quantum corrections were introduced via the Coleman--Weinberg mechanism, and gravitational loop effects were incorporated to yield a renormalization-group (RG) improved description of the model. Our analysis revealed that the scale-dependent effective mass and coupling parameters can shift the phase boundaries between stable and unstable regimes, potentially preventing collapse in parameter spaces where the classical theory might otherwise predict instability.

Focusing on the non-relativistic limit, we derived the coupled Gross--Pitaevskii--Poisson system that governs the dynamics of a gravitational exciton Bose--Einstein condensate (BEC). Numerical simulations of this system demonstrated that the condensate relaxes into a stable, low-energy state characterized by a diffuse density profile and a shallow gravitational potential well. The conservation of the mean density during the evolution confirmed particle-number conservation, while the gradual smoothing of density fluctuations and the flattening of the gravitational potential corroborated the expected long-term behavior of a self-gravitating BEC.

Our stability analysis proceeded by seeking stationary solutions via imaginary time evolution. In the spherically symmetric case, we obtained a ground state with a chemical potential \(\mu \approx 0.05057\) and an energy \(E \approx 0.00587\), with the density peaking at approximately 0.00336 and a corresponding shallow gravitational potential minimum of about \(-0.00588\). These results underscore that the condensate, while exhibiting a central density enhancement, is overall diffuse—a feature that is consistent with the expected behavior of dark matter candidates.

Furthermore, our Bogoliubov analysis of small fluctuations about a homogeneous condensate revealed that the excitation spectrum follows the well-known dispersion relation
\[
\hbar^2\omega^2 = \epsilon_k\Bigl(\epsilon_k+2\lambda_{\mathrm{eff}}^{\rm NR}n_0\Bigr),
\]
with the free-particle energy \(\epsilon_k = \hbar^2k^2/(2m_{\rm eff})\). In the long-wavelength limit, the dispersion relation becomes linear, \(\omega\approx c_s\,k\), where the speed of sound is given by \(c_s = \sqrt{\lambda_{\mathrm{eff}}^{\rm NR}n_0/m_{\rm eff}}\). The numerical results, which yield \(c_s \approx 6.1237\times10^{-2}\), validate the emergence of phonon-like excitations in the condensate. We further discussed how finite-temperature effects, nonlinear excitations (such as solitons and shock waves), and coupling to gravitational waves might modify the Bogoliubov spectrum, potentially offering novel observational signatures in cosmic microwave background anisotropies and the dynamics of dwarf galaxies.

Overall, our study demonstrates that gravitational exciton condensates possess a rich and robust phenomenology, supported by both analytical derivations and numerical simulations. The interplay between extra-dimensional stabilization, quantum corrections, and nonlinear self-interactions not only provides a viable mechanism for stabilizing the internal dimensions and generating an effective cosmological constant but also suggests that these condensates may serve as promising dark matter candidates. The diffuse density profiles, long lifetimes, and weak interactions of gravitational excitons align well with astrophysical observations, while the predicted Bogoliubov excitations and their modified dispersion relations open up new avenues for indirect detection via gravitational wave observations.

Future work will extend this analysis to non-spherically symmetric and rotating configurations, investigate finite-temperature and damping effects in greater detail, and explore the formation of topological solitons. Such studies will further explain the role of gravitational exciton condensates in the evolution of cosmic structures and contribute to our understanding of the dark sector of the Universe.

\bibliography{apssamp}

\end{document}